\documentclass[
reprint,
dcolumn,
bm,
nofootinbib,
notitlepage,
superscriptaddress]{revtex4-2}
\usepackage{lmodern}
\usepackage[latin1]{inputenc}
\usepackage[T1]{fontenc}
\usepackage{graphicx}
\usepackage{xcolor}
\usepackage{amsmath}
\usepackage{amssymb}
\usepackage{braket}
\usepackage{mathrsfs}
\usepackage{mathtools}
\usepackage{hyperref}
\usepackage{ulem}
\usepackage{upgreek}
\usepackage{pstricks-add}
\usepackage{natbib} 
\usepackage{listings}
\usepackage{tikz}
\usepackage{tcolorbox}
\usetikzlibrary{mindmap}
\usepackage{cancel}
\usepackage{algorithm2e}
\usepackage{simpler-wick}
\usepackage{pifont}
\definecolor{lightblue}{rgb}{0.145,0.6666,1}

\begin{document}
\title{Cavity-Modified Zeeman Effect via Spin-Polariton Formation}

\author{Eric W. Fischer}
\email{eric.fischer@hu-berlin.de}
\affiliation{Humboldt-Universit\"at zu Berlin, Institut f\"ur Chemie, Brook-Taylor-Stra\ss e 2, D-12489 Berlin, Germany}

\author{Michael Roemelt}
\affiliation{Humboldt-Universit\"at zu Berlin, Institut f\"ur Chemie, Brook-Taylor-Stra\ss e 2, D-12489 Berlin, Germany}

\date{\today}

\let\newpage\relax

\begin{abstract}
We study the electronic spin Zeeman effect for an effective spin-$1/2$-system subject to both strong coupling to a low-frequency optical cavity and an external static magnetic field. Specifically, we address the interplay between the cavity magnetic field component in a cavity Zeeman interaction and the canonical spin Zeeman interaction from the perspective of an effective spin-polariton Hamiltonian. The latter is derived from the minimal coupling Pauli-Fierz Hamiltonian beyond the common dipole approximation via first-order quasi-degenerate perturbation theory. We find the spin Zeeman effect to be modified in the presence of the cavity field due to the formation of spin-polariton states, which result from an intricate interplay of cavity and external magnetic fields in our model. Spin-polariton signatures are discussed in the context of electron paramagnetic resonance (EPR) spectroscopy along with cavity-induced modifications of the electronic g-factor. 
\end{abstract}

\let\newpage\relax
\maketitle
\newpage

\section{Introduction}
\label{sec.intro}
The experimental realization of strong light-matter coupling with molecules sparked the rapidly evolving field of polaritonic chemistry conceptually located at the intersection of quantum optics and molecular chemistry.\cite{ebbesen2016} In polaritonic chemistry, the quantum character of confined electromagnetic modes in optical microcavities is exploited to modify molecular properties and chemical reactivity.\cite{ebbesen2016,garciavidal2021} By tuning those cavity modes resonant to either electronic or vibrational excitations of molecules, the corresponding electronic strong coupling (ESC) and vibrational strong coupling (VSC) regimes have been experimentally characterized to exhibit intriguingly altered excited and ground state reactivity.\cite{hutchison2012,thomas2016} Moreover, the effect of a confined cavity field on magnetic and spin properties of molecules been experimentally investigated in the context of molecular magnets\cite{eddins2014}, quantum computation\cite{jenkins2016,bonizzoni2017} and very recently nuclear magnetic resonance (NMR) spectroscopy\cite{patrahau2024}. 

From a theoretical perspective, \textit{ab initio} approaches to polaritonic chemistry have been introduced for both ESC and VSC regimes via quantum electrodynamics (QED) extensions of quantum chemical wave function methods\cite{haugland2020,riso2022} and density functional theory\cite{tokatly2013,ruggenthaler2014}, besides cavity Born-Oppenheimer wave function approaches\cite{schnappinger2023,fischer2024,fischer2025}. In contrast, magnetic properties in the strong coupling regime are significantly less studied theoretically. Only recently, Rokaj \textit{et al.}\cite{rokaj2019,rokaj2022} investigated solid state systems subject to cavity strong coupling in presence of an external static magnetic field. Subsequently, magnetic properties of molecular systems under strong coupling were discussed by Barlini \textit{et al.}\cite{barlini2024}, who exploited QED Hartree-Fock theory to address cavity-modified molecular aromaticity taking into account both external static and nuclear magnetic effects. However, in both cases the cavity magnetic field component was eventually not taken into account.

In this study, we aim at extending these results by investigating electronic spins in presence of \textit{both} an external static magnetic field \textit{and} the quantized cavity magnetic field component of a low-frequency cavity. To this end, we derive an effective spin-polariton Hamiltonian for a molecular system with a single unpaired electron from the minimal coupling Pauli-Fierz Hamiltonian with both classical and cavity Zeeman interactions. Methodologically, we exploit an effective Hamiltonian approach combined with quasi-degenerate perturbation theory well established in the context of effective spin Hamiltonians employed in quantum chemistry.\cite{mcweeny1965,neese1998,neese2017} The cavity Zeeman interaction is treated from an approximate ``beyond-dipole'' perspective in line with similar initial ideas discussed in Ref.\cite{barlini2024}. We report on cavity-induced modifications of the spin Zeeman effect resulting from the formation of spin-polariton states in the presence of low-frequency cavity modes. Inspired by recent NMR experiments in the VSC regime\cite{patrahau2024}, we discuss our results in the context of electron paramagnetic resonance (EPR) spectroscopy and extract cavity-induced modifications of the electronic g-factor from the effective spin-polariton approach.

\section{Theory}
\subsection{Effective Spin-Polariton Hamiltonian}
\label{sec.cavity_spin_hamilton}
We study the interaction of an electronic spin system with effective spin, $S=1/2$, and quantized field modes of a low-frequency Fabry-P\'erot cavity from the perspective of an effective spin-polariton Hamiltonian
\begin{align}
\hat{H}_\mathrm{eff}
=
\hat{H}_\mathrm{Zee}
+
\hat{H}_\mathrm{cZee}
+
\hat{H}_c
\quad.
\label{eq.spin_polariton_hamilton}
\end{align}
We derive $\hat{H}_\mathrm{eff}$ from the Pauli-Fierz Hamiltonian in minimal coupling form via first-order quasi-degenerate perturbation theory while accounting for leading-order beyond-dipole contributions (\textit{cf.} Appendix \ref{sec.derivation}). The first term of Eq.\eqref{eq.spin_polariton_hamilton} resembles the canonical spin Zeeman Hamiltonian, which we consider for a classical B-field aligned parallel to the $z$-axis 
\begin{align}
\hat{H}_\mathrm{Zee}
&=
\dfrac{g_e\mu_B}{\hbar}
\hat{S}_z
B_z
\quad,
\label{eq.canonical_zeeman_hamilton}
\end{align}
with electronic g-factor of the free electron, $g_e$, Bohr magneton, $\mu_B$, and effective spin-$1/2$-operator, $\hat{S}_z=\frac{\hbar}{2}\sigma_z$, where $\sigma_z$ is the Pauli-Z-matrix. The second term in Eq.\eqref{eq.spin_polariton_hamilton} constitutes the cavity Zeeman Hamiltonian 
\begin{align}
\hat{H}_\mathrm{cZee}
&=
\dfrac{g_e\mu_B}{\hbar}
\underline{\hat{S}}
\cdot
\underline{\hat{B}}_c
\quad,
\label{eq.cavity_zeeman_hamilton}
\end{align}
with $\underline{\hat{S}}=(\hat{S}_x,\hat{S}_y,\hat{S}_z)^T$, where we treat the cavity magnetic-field operator, $\hat{B}_c$, approximately as
\begin{align}
\underline{\hat{B}}_c
&=
\mathrm{i}
g_0
\sqrt{\dfrac{\hbar}{2\omega_c}}
\sum^2_{\lambda=1}
(
\underline{k}
\times
\underline{e}_\lambda
)
\left(
\hat{b}_\lambda
-
\hat{b}^\dagger_\lambda
\right)
\quad,
\label{eq.approx_cavity_b_operator}
\end{align}
following recent ideas discussed in Ref.\cite{barlini2024} (\textit{cf.} Appendix \ref{sec.derivation} for details). In Eq.\eqref{eq.approx_cavity_b_operator}, we have the light-matter interaction constant, $g_0$, which will here be treated as free parameter and a vector of Pauli matrices, $\underline{\sigma}=(\sigma_x,\sigma_y,\sigma_z)^T$. Moreover, we consider an effective cavity mode with harmonic frequency, $\omega_c$, wavevector, $\underline{k}$, and creation(annihilation) operators, $\hat{b}^\dagger_\lambda(\hat{b}_\lambda)$, which is doubly degenerate with respect to two orthogonal polarization directions, $\lambda$, with related polarization vectors, $\underline{e}_\lambda$, respectively. Finally, the cavity Hamiltonian is given by
\begin{align}
\hat{H}_c
&=
\hbar\omega_c
\sum^2_{\lambda=1}
\hat{b}^\dagger_\lambda
\hat{b}_\lambda
\quad,
\end{align}
where we omit the zero-point energy, which constitutes just a total energy shift.

\subsection{Spin-Cavity-Interaction Scenarios}
For a classical static $B_z$-field, the eigenvalues of $\hat{H}_\mathrm{Zee}$ differ by the Zeeman splitting
\begin{align}
\Delta_\mathrm{Zee}
=
g_e
\mu_B
B_z
\quad.
\end{align} 
Herein, we are interested in how the presence of the cavity field modifies $\Delta_\mathrm{Zee}$ and consequently the electronic g-factor, $g_e$, as a function of the light-matter interaction strength, $g_0$. To this end, we shall first discuss the possible spin-cavity interaction scenarios. Orthogonal vectors, $\{\underline{k}_x,\underline{e}_y,\underline{e}_z\}$, span a reference coordinate frame, the cavity frame, which we align such that the wavevector is parallel to the $x$-axis, while the two polarization vectors span the $y$-$z$-plane.

Due to the cross product in Eq.\eqref{eq.approx_cavity_b_operator}, the cavity B-field operator takes two different forms depending on the cavity mode polarization
\begin{align}
\lambda
=
z
:
&
\quad
\underline{u}_x
\times
\underline{e}_z
=
\underline{e}_y
\quad,
\label{eq.z_polar_cavity}
\vspace{0.2cm}
\\
\lambda
=
y
:
&
\quad
\underline{u}_x
\times
\underline{e}_y
=
\underline{e}_z
\quad,
\label{eq.y_polar_cavity}
\end{align}
where we introduced a unit vector, $\underline{u}_x$, along the $x$-axis which allows us to write, $\underline{k}_x=\frac{\omega_c}{c}\underline{u}_x$, with speed of light, $c$, respectively. In the first scenario, Eq.\eqref{eq.z_polar_cavity}, the cavity B-field acts along the y-axis of the cavity frame orthogonal to the classical B-field along the z-axis. The second scenario is characterized by both the cavity and the classical B-field being oriented along the z-axis (\textit{cf.} Eq.\eqref{eq.y_polar_cavity}). 
In order to disentangle the two interaction scenarios, we will first discuss them separately in an effective single-mode-limit (\textit{cf.} Sec.\ref{eq.cavity_zeeman_effect}). This analysis already provides access to the main results of this study, as will be seen from a subsequent discussion of the full two-mode scenario (\textit{cf.} Sec.\ref{sec.two_modes}). 

In the single-mode-limit for a mode with polarization $\lambda$, we represent the spin-polariton Hamiltonian in a zero-order product basis
\begin{align}
\ket{\uparrow,0_\lambda},
\ket{\downarrow,0_\lambda},
\ket{\uparrow,1_\lambda},
\ket{\downarrow,1_\lambda}
\quad,
\label{eq.product_basis}
\end{align}
which is composed of a Kramers doublet ($\ket{\uparrow},\ket{\downarrow}$) and two cavity states ($\ket{0_\lambda},\ket{1_\lambda}$), as we restrict our discussion to a single cavity photon. A $z$-polarized cavity mode (\textit{cf.} Eq.\eqref{eq.z_polar_cavity}) turns the cavity Zeeman operator into
\begin{align}
\hat{H}_\mathrm{cZee}
=
\mathrm{i}
\dfrac{g_0\,g_e\mu_B}{2c}
\sqrt{\dfrac{\hbar\omega_c}{2}}
\sigma_y
\left(
\hat{b}_z
-
\hat{b}^\dagger_z
\right)
\quad,
\label{eq.zpol_cavity_zeeman_hamilton}
\end{align}
which contains the Pauli-Y-matrix, $\sigma_y$, and leads to a block-diagonal spin-polariton Hamiltonian
\begin{align}
\underline{\underline{H}}_\mathrm{eff}
&=
\begin{pmatrix}
\underline{\underline{H}}_p & 0
\vspace{0.2cm}
\\
0         & \underline{\underline{H}}_s
\end{pmatrix}
\quad.
\label{eq.zpol_spin_polariton_hamilton}
\end{align}
The upper diagonal entry reads
\begin{align}
\underline{\underline{H}}_p
&=
\begin{pmatrix}
\dfrac{g_e\mu_B}{2}
B_z
&
\dfrac{g_0\,g_e\mu_B}{2c}
\sqrt{\dfrac{\hbar\omega_c}{2}}
\vspace{0.2cm}
\\
\dfrac{g_0\,g_e\mu_B}{2c}
\sqrt{\dfrac{\hbar\omega_c}{2}}
&
-
\dfrac{g_e\mu_B}{2}
B_z
+
\hbar\omega_c
\end{pmatrix}
\quad,
\label{eq.polariton_hamilton}
\end{align}
and is spanned by states, $\ket{\uparrow,0_z},\ket{\downarrow,1_z}$, which come energetically close when $B_z$ increases. This facilitates light-matter interaction mediated via off-diagonal matrix elements subsequently leading to spin-polariton formation (indicated by index $p$) (\textit{cf.} Sec.\ref{subsec.spin_polariton}). We like to note that spin-polariton formation results here from the interplay of both classical and cavity magnetic fields. Moreover, the lower-diagonal component in Eq.\eqref{eq.zpol_spin_polariton_hamilton} reads 
\begin{align}
\underline{\underline{H}}_s
&=
\begin{pmatrix}
\dfrac{g_e\mu_B}{2}
B_z
+
\hbar\omega_c
&
-
\dfrac{g_0\,g_e\mu_B}{2c}
\sqrt{\dfrac{\hbar\omega_c}{2}}
\vspace{0.2cm}
\\
-
\dfrac{g_0\,g_e\mu_B}{2c}
\sqrt{\dfrac{\hbar\omega_c}{2}}
&
-
\dfrac{g_e\mu_B}{2}
B_z
\end{pmatrix}
\quad,
\label{eq.spectator_hamilton}
\end{align}
and is spanned by states, $\ket{\downarrow,0_z},\ket{\uparrow,1_z}$, respectively. In contrast to $\underline{\underline{H}}_p$, the energy difference between basis states spanning $\underline{\underline{H}}_s$ increases with $B_z$, such that they do not contribute to spin-polariton formation and are accordingly denoted as \textit{spectator} states (index $s$). However, the lower lying spectator state resembles the ground state of the full spin-polariton Hamiltonian, which becomes relevant for extracting the cavity-modified Zeeman splitting, $\tilde{\Delta}_\mathrm{zee}$, as discussed in Sec.\ref{subsec.cavity_zeeman_split}. 

Eventually, the second interaction scenario related to a $y$-polarized cavity mode (\textit{cf.} Eq.\eqref{eq.y_polar_cavity}) does not exhibit spin-polariton states but will become relevant for the two-mode scenario discussed in Sec.\ref{sec.two_modes}. The absence of spin-polariton states is understood by noting that a $y$-polarized cavity mode leads to a cavity Zeeman interaction, which does not couple different spin sectors but identical spins exhibiting the same interaction with $B_z$, \textit{i.e.}, both states are either raised or lowered in energy (\textit{cf.} Appendix \ref{sec.no_polariton_scenario} for details).  
\begin{figure*}[hbt]
\begin{center}
\includegraphics[scale=1.0]{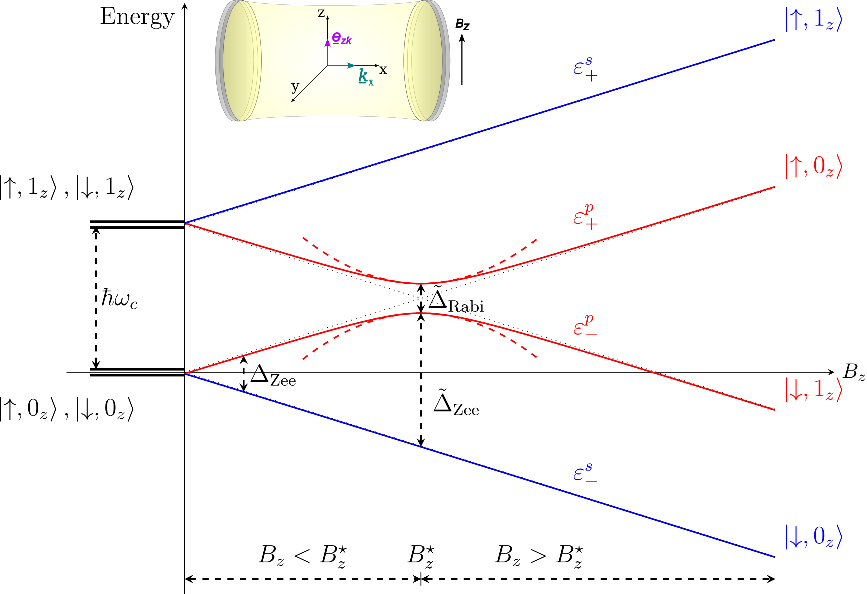}
\end{center}
\renewcommand{\baselinestretch}{1.}
\caption{Eigenvalues of the spin-polariton Hamiltonian for a single cavity mode polarized along the $z$-axis as function of static B-field strength, $B_z$. Spin-polariton branches related to eigenvalues $\varepsilon^p_\mp$ in Eq.\eqref{eq.spin_polariton_energies} (colored in red) and spectator state branches related to eigenvalues $\varepsilon^s_\mp$ in Eq.\eqref{eq.spectator_energies} (colored in blue). Additionally shown is the canonical Zeeman splitting, $\Delta_\mathrm{Zee}$, besides its cavity-modified counterpart, $\tilde{\Delta}_\mathrm{Zee}$, and the spin-polariton Rabi splitting, $\tilde{\Delta}_\mathrm{Rabi}$; the latter two for the resonance static B-field value $B^\star_z$ given in Eq.\eqref{eq.res_B_field}. The inset shows a sketch of the cavity frame for the single-mode scenario and the orientation of an external static B-field.}
\label{fig.cavity_zeeman}
\end{figure*}

\section{Cavity-Modified Zeeman Effect}
\label{eq.cavity_zeeman_effect}
We now turn to a detailed analysis of the spin-polariton Hamiltonian for a single $z$-polarized cavity mode given by Eq.\eqref{eq.zpol_spin_polariton_hamilton}. In Fig.\ref{fig.cavity_zeeman}, we qualitatively show the related eigenenergies as a function of the classical B-field amplitude, $B_z$, at fixed light-matter interaction, $g_0$. The excitation energy of zero-order states spanning $\underline{\underline{H}}_p$ is given by $\Delta_p=\hbar\omega_c-g_e\mu_B B_z$, and we discuss spin-polariton formation in the resonance scenario, $\Delta_p=0$. In this case, the classical B-field amplitude reads
\begin{align}
B^\star_z
=
\dfrac{\hbar\omega_c}{g_e \mu_B}
\quad,
\label{eq.res_B_field}
\end{align}
and is solely characterized by the cavity frequency. For an EPR experiment operating at $94\,\mathrm{GHz}$, which corresponds to a magnetic field strength of $B^\star_z=3.35\,\mathrm{T}$, one would therefore require a cavity frequency, $\hbar\omega_c=3.1\,\mathrm{cm}^{-1}$. At this point, we like to note that the effective spin-polariton Hamiltonian in Eq.\eqref{eq.spin_polariton_hamilton} fully neglects canonical zero-field splitting (ZFS) effects, that could arise due to spin-orbit coupling (SOC) or spin-spin coupling present also in absence of an external static magnetic field. In principle, such ZFS can affect the resonance scenario. Moreover, we do not take into account cavity-induced corrections of SOC for example, which might potentially contribute to spin-polariton formation in addition to the cavity Zeeman term.

\subsection{Spin-Polariton Formation}
\label{subsec.spin_polariton}
The eigenvalues of $\underline{\underline{H}}_p$ in Eq.\eqref{eq.polariton_hamilton} are generally given by
\begin{align}
\varepsilon^{p}_\mp
&=
\dfrac{\hbar\omega_c}{2}
\mp
\dfrac{
\sqrt{
(
g_e
\mu_B
B_z
-
\hbar\omega_c
)^2
+
g^2_0
\dfrac{g^2_e\mu^2_B}{2c^2}
\hbar\omega_c
}
}{2}
\,.
\label{eq.spin_polariton_energies}
\end{align}
Fig.\ref{fig.cavity_zeeman} depicts $\varepsilon^{p}_\mp$ as a function of $B_z$. In the resonance scenario, we obtain lower ($-$) and upper ($+$) spin-polariton states
\begin{align}
\ket{\mp}
&=
\dfrac{1}{\sqrt{2}}
\left(
\ket{\uparrow,0_z}
\mp
\ket{\downarrow,1_z}
\right)
\end{align}
with eigenenergies 
\begin{align}
\varepsilon^{p}_\mp
&=
\dfrac{\hbar\omega_c}{2}
\mp
\dfrac{g_0 g_e\mu_B}{2c}
\sqrt{
\dfrac{\hbar\omega_c}{2}
}
\quad,\,
B_z
=
B^\star_z
\quad.
\label{eq.spin_polariton_energies_res}
\end{align}
The corresponding Rabi splitting 
\begin{align}
\tilde{\Delta}_\mathrm{Rabi}
&=
\dfrac{g_0 g_e\mu_B}{c}
\sqrt{
\dfrac{\hbar\omega_c}{2}
}
\quad,
\label{eq.spin_polariton_rabi}
\end{align}
leads to a cavity-modified Zeeman splitting as discussed in Sec.\ref{subsec.cavity_zeeman_split}. We can now introduce an effective coupling constant
\begin{align}
\eta
&=
\dfrac{\tilde{\Delta}_\mathrm{Rabi}}{2\hbar\omega_c}
=
\dfrac{g_0 g_e\mu_B}{2c\sqrt{2\hbar\omega_c}}
\quad,
\end{align}
which defines the strong coupling regime via experimentally accessible parameters as $\eta\leq0.1$. Note, a proper lower bound of $\eta$ would require to account for dissipation channels in the light-matter hybrid system, which we neglect here for the sake of simplicity.

In the vicinity of $B^\star_z$, we furthermore observe a quadratic dependence of spin-polariton energies on the classical B-field amplitude
\begin{align}
\varepsilon^p_\mp
\approx
\varepsilon^p_\mp(B^\star_z)
\mp
\dfrac{g_e\mu_B c}{\sqrt{4g^2_0\hbar\omega_c}}
(\Delta B_z)^2
\quad,\,
g_0
>
0
\quad,
\label{eq.spin_polariton_curvature}
\end{align} 
with $\Delta B_z=B_z-B^\star_z$ as indicated by dashed parabolas in Fig.\ref{fig.cavity_zeeman}. 

We shall now consider classical B-field amplitudes distinct from the resonance scenario. In the B-field-free limit, $B_z\to0$, basis states $\ket{\uparrow,0_z}$ and $\ket{\downarrow,1_z}$ are approximately separated by the cavity mode excitation energy, $\hbar\omega_c$, which prevents spin-polariton formation. However, when we inspect the approximate eigenvalues
\begin{align}
\varepsilon^p_\mp
&\approx
\dfrac{\hbar\omega_c}{2}
\mp
\left(
\dfrac{\hbar\omega_c}{2}
+
g^2_0
\dfrac{g^2_e\mu^2_B}{8c^2}
\right)
\quad,
\label{eq.spin_polariton_energies_weak}
\end{align}
which we obtain via a Taylor expansion of Eq.\eqref{eq.spin_polariton_energies} around $B_z=0$, we observe a slight repulsion of both states due to non-zero light-matter interaction, $g_0$, with energy difference
\begin{align}
\Delta_0
&=
\hbar\omega_c
+
g^2_0
\dfrac{g^2_e\mu^2_B}{4c^2}
\quad.
\end{align}
We note here that the second term is quadratic in the light-matter interaction constant, $g_0$, in contrast to the Rabi splitting, $\tilde{\Delta}_\mathrm{Rabi}$, which is linear in $g_0$. Eventually, in the strong B-field regime, one finds 
\begin{align}
\varepsilon^p_\mp
&\approx
\dfrac{\hbar\omega_c}{2}
\mp
\dfrac{
g_e
\mu_B
B_z
}{2}
\quad,\,
B_z
> 
B^\star_z
\quad,
\end{align}
with an energy difference fully determined by the classical B-field amplitude
\begin{align}
\Delta_\mathrm{strong}
&=
\Delta_\mathrm{Zee}
\quad,
\end{align}
equivalent to the classical Zeeman splitting. Here, we like to point out two subtleties in context of the strong B-field regime: first, this notion is only meaningful as long as formation of energetically higher lying spin-polariton states can be neglected. Such states will emerge when one additionally includes the two-photon manifold (or even higher photon numbers), which we neglected in our model. Second, in case we can neglect latter interactions, $\varepsilon^p_-$ corresponds to $\ket{\downarrow,1_c}$ and $\varepsilon^p_+$ to $\ket{\uparrow,0_z}$, such that the energy, $\Delta_\mathrm{Zee}$, would be required to excite \textit{both} spin \textit{and} cavity subsystems, respectively, and does not only refer to a spin flip. 
\begin{figure*}[hbt!]
\begin{center}
\includegraphics[scale=1.0]{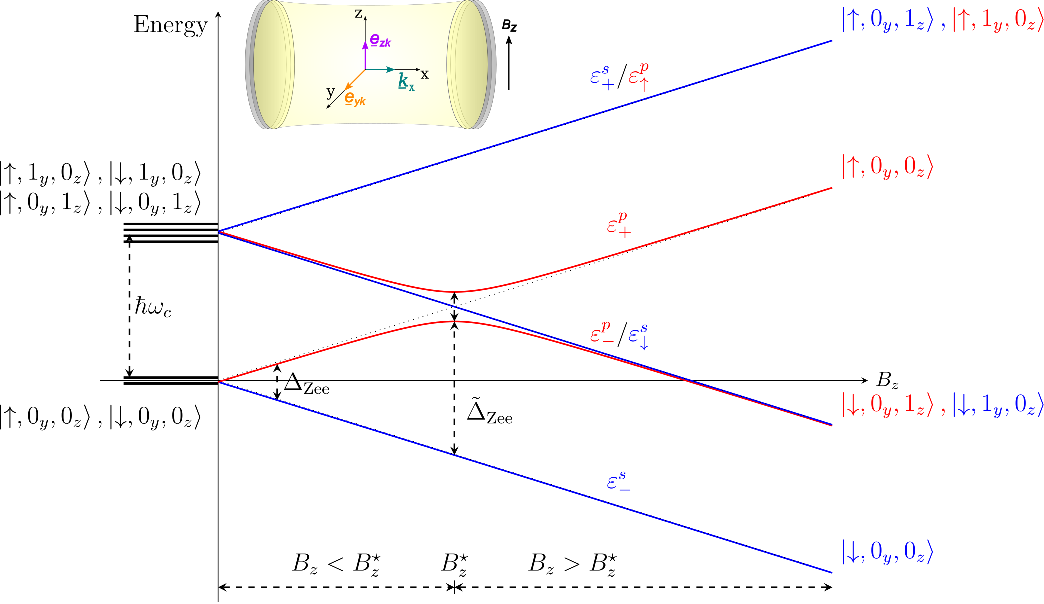}
\end{center}
\renewcommand{\baselinestretch}{1.}
\caption{Eigenvalues of the spin-polariton Hamiltonian for two cavity modes polarized along $y$- and $z$-axis as function of static B-field strength, $B_z$. Spin-polariton branches (colored in red) and spectator state branches (colored in blue) for the two-mode scenario described by the spin-polariton Hamiltonian in Eq.\eqref{eq.two_modes_spin_polariton_hamilton}. In comparison to the single-mode scenario presented in Fig.\ref{fig.cavity_zeeman}, two new branches appear with energies, $\varepsilon^s_\downarrow$ and $\varepsilon^p_\uparrow$, which correspond to unaltered zero-order product states, $\ket{\downarrow,1_y,0_z}$ and $\ket{\uparrow,1_y,0_z}$, containing one photon in the $y$-polarized cavity mode (\textit{cf.} Appendix \ref{sec.no_polariton_scenario}). The inset shows a sketch of the cavity frame for the two-mode scenario and the orientation of an external static B-field.}
\label{fig.cavity_zeeman_two_mode}
\end{figure*}

\subsection{Cavity-Modified Zeeman-Splitting}
\label{subsec.cavity_zeeman_split}
Eigenstates of $\underline{\underline{H}}_s$ in Eq.\eqref{eq.spectator_hamilton}, \textit{i.e.}, spectator states, are not involved in spin-polariton formation but contain the ground state of the spin-polariton Hamiltonian as observable from Fig.\ref{fig.cavity_zeeman}. The corresponding energies are generally given by
\begin{align}
\varepsilon^s_\mp
&=
\dfrac{\hbar\omega_c}{2}
\mp
\dfrac{
\sqrt{
(
g_e
\mu_B
B_z
+
\hbar\omega_c
)^2
+
g^2_0
\dfrac{g^2_e\mu^2_B}{2c^2}
\hbar\omega_c
}
}{2}
\quad,
\label{eq.spectator_energies}
\end{align}
and the ground state energy, $\varepsilon^s_-$, of the full spin-polariton Hamiltonian enters the cavity-modified Zeeman splitting
\begin{align}
\tilde{\Delta}_\mathrm{Zee}
&=
\varepsilon^p_-
-
\varepsilon^s_-
\quad.
\label{eq.cavity_zeeman_split_general}
\end{align}
In the resonance scenario, we find for the ground state energy
\begin{align}
\varepsilon^s_-
&\approx
-
\dfrac{\hbar\omega_c}{2}
+
\mathcal{O}(g^2_0)
\quad,\,
B_z
=
B^\star_z
\quad,
\label{eq.spectator_energies_resonance}
\end{align}
with a cavity-induced correction that is in leading order quadratic with respect to the light-matter interaction strength, $g_0$. Since, the lower spin-polariton energy, $\varepsilon^p_-$, in Eq.\eqref{eq.spin_polariton_energies_res} scales only linear with $g_0$, we can neglect this correction here to obtain the cavity-modified Zeeman splitting in the resonance scenario
\begin{align}
\tilde{\Delta}_\mathrm{Zee}
&=
\hbar\omega_c
-
\dfrac{g_0 g_e\mu_B}{2c}
\sqrt{
\dfrac{\hbar\omega_c}{2}
}
\quad,\,
B_z
=
B^\star_z
\quad.
\label{eq.cavity_zeeman_split_res}
\end{align}
From this result, we can now extract the cavity-modified electronic g-factor, $\tilde{g}_e$, as (\textit{cf.} Appendix \ref{sec.derivation_cavity_g_factor})
\begin{align}
\tilde{g}_e
&=
\dfrac{\tilde{\Delta}_\mathrm{Zee}}{\mu_B B^\star_z}
\left(
\dfrac{1}{1-\dfrac{g_0}{2B^\star_z}\sqrt{\dfrac{\hbar\omega_c}{2}}}
\right)
\quad,
\label{eq.cavity_modified_g_factor}
\end{align}
where we recover the bare electronic g-factor $\tilde{g}_e\to g_e$ via $\tilde{\Delta}_\mathrm{Zee}\to\Delta_\mathrm{Zee}$ in the non-interacting limit, $g_0\to 0$.

Conceptually, one could determine $\tilde{g}_e$ in the spirit of a conventional EPR experiment by measuring the cavity-modified Zeeman-splitting, $\tilde{\Delta}_\mathrm{Zee}$, spectroscopically. There one resonantly excites the light-matter hybrid system, $\tilde{\Delta}_\mathrm{Zee}=\hbar\omega_\mathrm{las}$, with an external resonant laser field of frequency, $\omega_\mathrm{las}$, as
\begin{align}
\ket{\downarrow,0_z}
\overset{\hbar\omega_\mathrm{las}}{\longrightarrow}
\dfrac{1}{\sqrt{2}}
\left(
\ket{\downarrow,1_z}
-
\ket{\uparrow,0_z}
\right)
\quad.
\end{align}
By inspecting the zero-order components of the lower spin-polariton state on the right-hand side, we realize that this can be achieved in two equivalent ways by either exciting the spin subsystem or its cavity counterpart. In the first case, one preserves the photon number while in the second case the spin state is preserved. 

Equivalently, one will also be able to excite the upper spin-polariton state
\begin{align}
\ket{\downarrow,0_z}
\overset{\hbar\omega_\mathrm{las}}{\longrightarrow}
\dfrac{1}{\sqrt{2}}
\left(
\ket{\downarrow,1_z}
+
\ket{\uparrow,0_z}
\right)
\quad,
\end{align}
by increasing the laser frequency, which leads to a doublet in a related EPR spectrum with energy spacing given by the Rabi splitting, $\tilde{\Delta}_\mathrm{Rabi}$, in Eq.\eqref{eq.spin_polariton_rabi}.

We close this section by addressing the B-field-free limit ($B_z\to 0$), which illustrates the character of light-matter interaction involving the cavity B-field from another perspective. In this case, the ground state energy is given by (\textit{cf.} Taylor expansion in Eq.\eqref{eq.spin_polariton_energies_weak})
\begin{align}
\varepsilon^s_-
&\approx
-
g^2_0
\dfrac{g^2_e\mu^2_B}{8c^2}
\quad,
\label{eq.spectator_energies_weak}
\end{align}
such that we find with Eq.\eqref{eq.spin_polariton_energies_weak} a \textit{vanishing} cavity-modified Zeeman splitting
\begin{align}
\tilde{\Delta}_\mathrm{Zee}
&=
0
\quad,\,
B_z\to 0
\quad,
\label{eq.cavity_zeeman_split_weak}
\end{align}
at non-zero light-matter interaction strength, $g_0$, respectively. This result is rationalized by recalling that the cavity B-field couples different spin and cavity mode excitation manifolds. However, Eq.\eqref{eq.cavity_zeeman_split_weak} relates to the energy difference between states $\ket{\downarrow,0_z}$ and $\ket{\uparrow,0_z}$, which are not coupled by the cavity B-field operator as they share an identical photon number. This analysis emphasizes the intricate interplay of both classical and cavity B-fields in the spin-polariton Hamiltonian. Moreover, canonical ZFS as well as related cavity-induced corrections, which we neglected at this stage of the analysis, might substantially alter spin-polariton formation.

\section{Two-Mode Scenario}
\label{sec.two_modes}
We finally discuss the two-mode scenario subject to a doubly-degenerate cavity mode polarized along $y$- and $z$-axis. In Sec.\ref{eq.cavity_zeeman_effect}, we have seen that a $z$-polarized cavity mode leads to a cavity Zeeman interaction with a $y$-polarized B-field component, which couples distinct spin sectors via excitation of a cavity photon. Moreover, we showed in Appendix \ref{sec.no_polariton_scenario}, that the cavity Zeeman interaction for a $y$-polarized mode contains a $z$-polarized cavity B-field preserving the spin states under excitation of a cavity photon. Based on those observations, we can partition the zero-order basis of the two-mode scenario now as
\begin{align}
\begin{matrix}
\ket{\uparrow,0_y,0_z},
\ket{\downarrow,0_y,1_z},
\ket{\uparrow,1_y,0_z}
\quad,
\vspace{0.2cm}
\\
\ket{\downarrow,0_y,0_z},
\ket{\uparrow,0_y,1_z},
\ket{\downarrow,1_y,0_z}
\quad,
\end{matrix}
\label{eq.product_basis_two_modes}
\end{align}
which leads again to a block-diagonal spin-polariton Hamiltonian 
\begin{align}
\underline{\underline{H}}^\prime_\mathrm{eff}
=
\begin{pmatrix}
\underline{\underline{H}}^\prime_p & 0
\vspace{0.2cm}
\\
0         & \underline{\underline{H}}^\prime_s
\end{pmatrix}
\quad,
\label{eq.two_modes_spin_polariton_hamilton}
\end{align}
with spin-polariton block
\begin{align}
\underline{\underline{H}}^\prime_p
=
\begin{pmatrix}
\dfrac{g_e\mu_B}{2}B_z
&
\mathrm{i}\,
b_0
&
b_0
\vspace{0.2cm}
\\
-
\mathrm{i}\,
b_0
&
\dfrac{g_e\mu_B}{2}
B_z
+
\hbar\omega_c
&
0
\vspace{0.2cm}
\\
b_0
&
0
&
-
\dfrac{g_e\mu_B}{2}
B_z
+
\hbar\omega_c
\end{pmatrix}
\,,
\end{align}
and spectator block
\begin{align}
\underline{\underline{H}}^\prime_s
=
\begin{pmatrix}
\dfrac{g_e\mu_B}{2}
B_z
+
\hbar\omega_c
&
-
b_0
&
0
\vspace{0.2cm}
\\
-
b_0
&
-
\dfrac{g_e\mu_B}{2}B_z
&
-
\mathrm{i}\,
b_0
\vspace{0.2cm}
\\
0
&
\mathrm{i}\,
b_0
&
-
\dfrac{g_e\mu_B}{2}
B_z
+
\hbar\omega_c
\end{pmatrix}
\,,
\end{align}
where we introduced 
\begin{align}
b_0
&=
\dfrac{g_0\,g_e\mu_B}{2c}
\sqrt{\dfrac{\hbar\omega_c}{2}}
\quad.
\end{align}
In Fig.\ref{fig.cavity_zeeman_two_mode}, we show the two sets of eigenvalues, $\varepsilon^p_i$ and $\varepsilon^s_i$ for $i=0,1,2$, as function of the static B-field amplitude, $B_z$, in analogy to the single-mode scenario discussed before. In addition to the two spin-polariton branches discussed in Fig.\ref{fig.cavity_zeeman}, we find another energetically close branch corresponding to the state, $\ket{\downarrow,1_y,0_z}$, which does not participate in spin-polariton formation. In a potential EPR experiment, which exploits excitation of the cavity instead of the spin subsystem, this branch leads to a third transition at an energy, $\hbar\omega_c$, besides the two spin-polariton peaks for $B_z=B^\star_z$. Notably, this is the only distinct two-mode feature, which has not yet been covered by our analysis of the simplified single-mode problem. 

\section{Conclusion}
\label{sec.conclusion}
We analysed cavity-induced modifications of the electronic spin Zeeman effect for an effective spin-$1/2$-system interacting with both an external static B-field and the quantized cavity B-field component. To this end, we derived an effective spin-polariton Hamiltonian from the minimal coupling Pauli-Fierz Hamiltonian accounting for beyond-dipole cavity field corrections by means of first-order quasi-degenerate perturbation theory. We discussed spin-polariton formation resulting from the interplay of both classical static and quantized cavity B-field components, while being subject to a Rabi splitting fully determined by the cavity field. The resonance scenario is characterized by a static B-field amplitude directly proportional to the cavity mode frequency and a cavity-modified Zeeman splitting smaller than the cavity-free scenario results from spin-polariton formation. We consequently derived a cavity-induced correction of the electronic g-factor and discussed a connection to potential EPR experiments based on relevant EPR spectroscopic signatures. Notably, we obtain all relevant features of spin-polariton formation already in the single-mode limit of our model with a cavity mode polarized parallel to the external static B-field. Another level of complexity can be added in the future by taking into account spin-orbit coupling and related cavity-induced corrections. Our study explores an intricate alternative light-matter interaction setting next to vibrational and electronic strong coupling scenarios, that hints at possibly non-trivial modifications of electronic spin properties in low-frequency optical cavities.


\section*{Acknowledgements}
E.W. Fischer acknowledges funding by the Deutsche Forschungsgemeinschaft (DFG, German Research Foundation) through DFG project 536826332.

\section*{Data Availability Statement}
The data that support the findings of this study are available from the corresponding author upon reasonable request.

\section*{Conflict of Interest}
The authors have no conflicts to disclose.

\renewcommand{\thesection}{}
\section*{Appendix}

\setcounter{equation}{0}
\renewcommand{\theequation}{\thesubsection.\arabic{equation}}
\subsection{Spin-Polariton Hamiltonian via Effective Hamiltonian Theory}
\label{sec.derivation}

\subsubsection{Pauli-Fierz Hamiltonian with Spin Zeeman Terms}
We consider the Pauli-Fierz Hamiltonian in minimal coupling form for electrons coupled to effective quantized field modes of a Fabry-P\'erot cavity
\begin{align}
\hat{H}
=
\sum^{N_e}_i
\dfrac{\underline{\hat{\pi}}^2_i}{2m_e}
+
V_\mathrm{cou}
+
\hat{H}_\mathrm{Zee}
+
\hat{H}_\mathrm{cZee}
+
\hat{H}_c
\quad,
\label{eq.minimal_coup_pauli_fierz_zeeman}
\end{align}
with molecular Coulomb interaction, $V_\mathrm{cou}$, canonical spin-Zeeman interaction, $\hat{H}_\mathrm{Zee}$, cavity Zeeman interaction, $\hat{H}_\mathrm{cZee}$, and cavity mode Hamiltonian, $\hat{H}_c$, respectively. The electronic kinetic momentum operators is given by
\begin{align}
\underline{\hat{\pi}}_i
&=
\underline{\hat{p}}_i
+
e
\underline{A}_\mathrm{cl}(\underline{r}_i)
+
e
\underline{\hat{A}}_c(\underline{r}_i,t)
\quad,
\label{eq.kin_momentum_electron}
\end{align}
with canonical momentum operator, $\underline{\hat{p}}_i$, and we work in Coulomb gauge, $\underline{\nabla}\cdot\underline{A}_\mathrm{cl}=0=\underline{\nabla}\cdot\underline{\hat{A}}_\mathrm{c}$, respectively. We consider an external static B-field along the $z$-axis resulting from a classical vector potential
\begin{align}
\underline{A}_\mathrm{cl}(\underline{r}_i)
&=
\dfrac{1}{2}\,
\underline{r}_i
\times
\underline{B}_\mathrm{cl}
\quad,
\quad
\underline{B}_\mathrm{cl}
=
(0,0,B_z)^T
\quad,
\label{eq.class_vecpot}
\end{align}
with classical B-field amplitude, $B_z$. The quantized cavity vector potential is given by
\small{
\begin{align}
\underline{\hat{A}}_c(\underline{r}_i,t)
=
g_0
\sum_{\lambda k}
\sqrt{\dfrac{\hbar}{2\omega_k}}
\underline{e}_{\lambda k}
(
\hat{b}_{\lambda k}(\underline{r}_i,t)
+
\hat{b}^\dagger_{\lambda k}(\underline{r}_i,t)
)
\quad,
\label{eq.cavity_vecpot}
\end{align}
}\normalsize
with light-matter interaction constant, $g_0$, harmonic cavity frequency, $\omega_k$, of the $k^\mathrm{th}$-cavity mode and cavity polarization vector, $\underline{e}_{\lambda k}$, for a mode with polarization, $\lambda$, respectively. Moreover, we have bosonic annihilation and creation operators
\begin{align}
\begin{matrix}
\hat{b}_{\lambda k}(\underline{r}_i,t)
&=
\hat{b}_{\lambda k}
e^{-\mathrm{i}\omega_k t}
e^{\mathrm{i}\underline{k}\cdot\underline{r}_i}
\vspace{0.2cm}
\\
\hat{b}^\dagger_{\lambda k}(\underline{r}_i,t)
&=
\hat{b}^\dagger_{\lambda k}
e^{\mathrm{i}\omega_k t}
e^{-\mathrm{i}\underline{k}\cdot\underline{r}_i}
\end{matrix}
\quad,
\label{eq.photon_operators}
\end{align}
with $[\hat{b}_{\lambda k},\hat{b}^\dagger_{\lambda^\prime k^\prime}]=\delta_{\lambda\lambda^\prime}\delta_{kk^\prime}$, and the zero-point energy shifted cavity field Hamiltonian reads
\begin{align}
\hat{H}_c
&=
\sum_{\lambda k}
\hbar\omega_k
\hat{b}^\dagger_{\lambda k}
\hat{b}_{\lambda k}
\quad.
\label{eq.cavity_field_hamiltonian}
\end{align}
The canonical spin Zeeman interaction follows via Eq.\eqref{eq.class_vecpot} as
\begin{align}
\hat{H}_\mathrm{Zee}
&=
\dfrac{g_e\mu_B}{\hbar}
\hat{S}_z
B_z
\quad,
\quad
\hat{S}_z
=
\dfrac{\hbar}{2}
\sum^{N_e}_i
\hat{\sigma}^i_z
\quad,
\label{eq.classical_zeeman}
\end{align}
with collective spin-1/2-operator, $\hat{S}_z$, determined by the Pauli-$Z$-matrix, $\hat{\sigma}_z$. The cavity Zeeman interaction is given by
\begin{align}
\hat{H}_\mathrm{cZee}
&=
\dfrac{g_e\mu_B}{\hbar}
\underline{\hat{S}}
\cdot
\underline{\hat{B}}_c
\quad,
\label{eq.cavity_zeeman}
\end{align}
with cavity magnetic field operator
\small{
\begin{align}
\underline{\hat{B}}_c
&=
\mathrm{i}
g_0
\sum_{\lambda k}
\sqrt{\dfrac{\hbar}{2\omega_k}}
(
\underline{k}
\times
\underline{e}_{\lambda k}
)
(
\hat{b}_{\lambda k}(\underline{r}_i,t)
-
\hat{b}^\dagger_{\lambda k}(\underline{r}_i,t)
)
\,.
\label{eq.cavity_b_field}
\end{align}
}\normalsize
We focus here solely on electronic spins, specifically, the spin of a single unpaired electron, and neglect nuclear spins.

\subsubsection{Beyond the Dipole Approximation}
We approximate coordinate-dependent photonic operators in Eq.\eqref{eq.photon_operators} as
\begin{align}
\begin{matrix}
\hat{b}_{\lambda k}(\underline{r}_i,t)
&\approx
\hat{b}_{\lambda k}(t)
(
1
+
\mathrm{i}\underline{k}\cdot\underline{r}_i
)
\vspace{0.2cm}
\\
\hat{b}^\dagger_{\lambda k}(\underline{r}_i,t)
&\approx
\hat{b}^\dagger_{\lambda k}(t)
(
1
-
\mathrm{i}\underline{k}\cdot\underline{r}_i
)
\end{matrix}
\quad,
\label{eq.beyond_dipole}
\end{align}
which resembles a first-order correction of the dipole approximation and leads to an approximate cavity vector potential
\begin{multline}
\underline{\hat{A}}_c(\underline{r}_i,t)
\approx
g_0
\sum_{\lambda k}
\sqrt{\dfrac{\hbar}{2\omega_k}}
\underline{e}_{\lambda k}
(
\hat{b}_{\lambda k}(t)
+
\hat{b}^\dagger_{\lambda k}(t)
)
\\
+
\mathrm{i}\,
g_0
\sum_{\lambda k}
\sqrt{\dfrac{\hbar}{2\omega_k}}
\underline{e}_{\lambda k}
(
\underline{k}\cdot\underline{r}_i
)
(
\hat{b}_{\lambda k}(t)
-
\hat{b}^\dagger_{\lambda k}(t)
)
\quad.
\label{eq.approx_cavity_vecpot}
\end{multline}
We obtain the corresponding approximate cavity electric field operator, $\underline{\hat{E}}_c=-\partial_t \underline{\hat{A}}_c$, as
\begin{multline}
\underline{\hat{E}}_c
\approx
\mathrm{i}\,
g_0
\sum_{\lambda k}
\sqrt{\dfrac{\hbar\omega_k}{2}}
\underline{e}_{\lambda k}
(
\hat{b}_{\lambda k}(t)
-
\hat{b}^\dagger_{\lambda k}(t)
)
\\
-
g_0
\sum_{\lambda k}
\sqrt{\dfrac{\hbar\omega_k}{2}}
\underline{e}_{\lambda k}
(
\underline{k}\cdot\underline{r}_i
)
(
\hat{b}_{\lambda k}(t)
+
\hat{b}^\dagger_{\lambda k}(t)
)
\quad,
\label{eq.approx_cavity_efield}
\end{multline}
and the approximate cavity magnetic field operator, $\underline{\hat{B}}_c=\underline{\nabla}\times\underline{\hat{A}}_c$, reads
\begin{align}
\underline{\hat{B}}_c
\approx
\mathrm{i}\,
g_0
\sum_{\lambda k}
\sqrt{\dfrac{\hbar}{2\omega_k}}
(
\underline{k}
\times
\underline{e}_{\lambda k}
)
(
\hat{b}_{\lambda k}(t)
-
\hat{b}^\dagger_{\lambda k}(t)
)
\,.
\label{eq.approx_cavity_bfield}
\end{align}
The herein derived approximate electric and magnetic field operator violate the Amp\`ere-Maxwell law in line with results discussed in Ref.\cite{barlini2024}.

\subsubsection{Effective Spin-Polariton Hamiltonian via Quasi-Degenerate Perturbation Theory}
We will now connect the Pauli-Fierz Hamiltonian in Eq.\eqref{eq.minimal_coup_pauli_fierz_zeeman} to the effective spin-polariton Hamiltonian in Eq.\eqref{eq.spin_polariton_hamilton}, respectively. To this end, we invoke an effective Hamiltonian approach in combination with (first-order) quasi-degenerate perturbation theory well established for Spin-Hamiltonian methods applied in quantum chemistry of transition metal complexes.\cite{mcweeny1965,neese1998} 

We start by partitioning the Pauli-Fierz Hamiltonian as
\begin{align}
\hat{H}
&=
\hat{H}_0
+
\hat{V}_1
\quad,
\end{align}
with
\begin{align}
\hat{H}_0
&=
\sum^{N_e}_i
\dfrac{\underline{\hat{\pi}}^2_i}{2m_e}
+
V_\mathrm{cou}
-
\hat{H}_\mathrm{oZee}
\quad,
\vspace{0.2cm}
\\
\hat{V}_1
&=
\hat{H}_\mathrm{oZee}
+
\hat{H}_\mathrm{Zee}
+
\hat{H}_\mathrm{cZee}
+
\hat{H}_c
\quad,
\end{align}
assuming cavity field operators in the approximation given by Eq.\eqref{eq.beyond_dipole}. Moreover, for the sake of completeness we also account for the orbital Zeeman term, $\hat{H}_\mathrm{oZee}$, in $\hat{V}_1$, which will however not be relevant for the final first-order result. We now note that the zeroth-order Pauli-Fierz Hamiltonian commutes with electronic spin operators
\begin{align}
[\hat{H}_0,\hat{S}^2]
&=
0
\quad,
\vspace{0.2cm}
\\
[\hat{H}_0,\hat{S}_z]
&=
0
\quad,
\end{align}
which allows us to label corresponding adiabatic states by spin and magnetic quantum numbers, $S$ and $M$, as
\begin{align}
\hat{H}_0
\ket{\tilde{\Psi}^{SM}_I}
&=
\tilde{E}_I
\ket{\tilde{\Psi}^{SM}_I}
\quad.
\label{eq.zero_order_pf_eigenvalue}
\end{align}
Those states mimic cavity-Born-Oppenheimer (CBO) adiabatic states introduced in the length-gauge description of polaritonic chemistry, which carry a parametric dependence on both nuclear and cavity displacement coordinates.\cite{flick2017,flick2017cbo,fischer2023} We may now write the full Pauli-Fierz Hamiltonian in the basis of adiabatic states as
\begin{align}
H^{JS^\prime M^\prime}_{ISM}
&=
\braket{
\tilde{\Psi}^{SM}_I
\vert
\hat{H}_0
+
\hat{V}_1
\vert
\tilde{\Psi}^{S^\prime M^\prime}_J
}
\quad,
\vspace{0.2cm}
\\
&=
\tilde{E}_I
\delta_{ISM,JS^\prime M^\prime}
+
\braket{
\tilde{\Psi}^{SM}_I
\vert
\hat{V}_1
\vert
\tilde{\Psi}^{S^\prime M^\prime}_J
}
\quad,
\end{align}
where the first term simplifies via Eq.\eqref{eq.zero_order_pf_eigenvalue}. We follow now Refs.\cite{mcweeny1965,neese1998} and assume a BO-scenario for the ground state multiplet and introduces a related $2M+1$-dimensional `a-set' of states, which couples to the remaining adiabatic states collected in a 'b-set' as
\begin{align}
\begin{pmatrix}
\underline{\underline{H}}_{aa} & \underline{\underline{H}}_{ab}
\\
\underline{\underline{H}}_{ba} & \underline{\underline{H}}_{bb}
\end{pmatrix}
\begin{pmatrix}
\underline{c}_a
\\
\underline{c}_b
\end{pmatrix}
&=
E
\begin{pmatrix}
\underline{c}_a
\\
\underline{c}_b
\end{pmatrix}
\quad.
\end{align}
After solving the latter formally for coefficient vectors, $\underline{c}_b$, one obtains an effective non-linear SE
\begin{align}
\left(
\underline{\underline{H}}_{aa}
-
\underline{\underline{H}}_{ab}
(
\underline{\underline{H}}_{bb}
-
E
\underline{\underline{1}}
)^{-1}
\underline{\underline{H}}_{ba}
\right)
\underline{c}_a
=
E
\underline{c}_a
\quad,
\end{align}
which is the simplified by invoking quasi-degenerate perturbation theory as follows: (i) One neglects coupling of b-states due to $\hat{V}_1$, which gives $H^{bb}_{KL}=E_K\delta_{KL}$ and (ii) one replaces $E\to\tilde{E}_0$ in the matrix inverse as one searches for states close to the ground state. One consequently obtains an effective Hamiltonian for the a-set, which reads up to second order
\begin{multline}
H^{MM^\prime}_\mathrm{eff}
=
\tilde{E}_0
\delta_{MM^\prime}
+
\braket{
\tilde{\Psi}^{SM}_0
\vert
\hat{V}_1
\vert
\tilde{\Psi}^{SM^\prime}_0}
\\
-
\sum_{K_bS^\prime M^{\prime\prime}}
\Delta^{-1}_{K_b}
\vert
\braket{
\tilde{\Psi}^{SM}_0
\vert
\hat{V}_1
\vert
\tilde{\Psi}^{S^\prime M^{\prime\prime}}_{K_b}}
\vert^2
\quad,
\end{multline}
with $\Delta_{K_b}=\tilde{E}_{K_b}-\tilde{E}_0$. From discussions of molecular spin Hamiltonians, we know that the orbital Zeeman term, $\hat{H}_\mathrm{oZee}$, in $\hat{V}_1$ does not contribute at first-order but appears only in a second-order correction.\cite{neese2017} We thus restrict the present discussion to first-order spin Zeeman effects captured by
\begin{align}
H^{MM^\prime}_\mathrm{eff}
&=
\braket{
\tilde{\Psi}^{SM}_0
\vert
\hat{H}_\mathrm{Zee}
+
\hat{H}_\mathrm{cZee}
+
\hat{H}_c
\vert
\tilde{\Psi}^{SM^\prime}_0}
\,,
\end{align}
which is an effective operator acting in spin-cavity space, where we concentrate here on $S=1/2$ and $M=\pm1/2$, respectively.

\setcounter{equation}{0}
\renewcommand{\theequation}{\thesubsection.\arabic{equation}}
\subsection{Spin-Polariton-Free Scenario}
\label{sec.no_polariton_scenario}
A $y$-polarized cavity mode ($\lambda=y$) leads to a cavity B-field along the $z$-axis parallel to its classical equivalent with cavity Zeeman Hamiltonian 
\begin{align}
\hat{H}_\mathrm{cZee}
=
\mathrm{i}
\dfrac{g_0\,g_e\mu_B}{2c}
\sqrt{\dfrac{\hbar\omega_c}{2}}
\sigma_z
\left(
\hat{b}_y
-
\hat{b}^\dagger_y
\right)
\quad.
\end{align}
In matrix representation (\textit{cf.} Eq.\eqref{eq.product_basis} with $\lambda=y$), the corresponding spin-polariton Hamiltonian is block diagonal 
\begin{align}
\underline{\underline{H}}_\mathrm{eff}
&=
\begin{pmatrix}
\underline{\underline{H}}_\uparrow
&
0
\vspace{0.2cm}
\\
0 
&
\underline{\underline{H}}_\downarrow
\end{pmatrix}
\quad,
\end{align}
with
\begin{align}
\underline{\underline{H}}_\uparrow
=
\begin{pmatrix}
\dfrac{g_e\mu_B}{2}B_z
&
\dfrac{\mathrm{i}\,g_0\,g_e\mu_B}{2c}
\sqrt{\dfrac{\hbar\omega_c}{2}}
\vspace{0.2cm}
\\
-
\dfrac{\mathrm{i}\,g_0\,g_e\mu_B}{2c}
\sqrt{\dfrac{\hbar\omega_c}{2}}
&
\dfrac{g_e\mu_B}{2}
B_z
+
\hbar\omega_c
\end{pmatrix}
\quad,
\end{align}
for states $\ket{\uparrow,0_y},\ket{\uparrow,1_y}$, and 
\begin{align}
\underline{\underline{H}}_\downarrow
=
\begin{pmatrix}
-
\dfrac{g_e\mu_B}{2}B_z
&
-
\dfrac{\mathrm{i}\,g_0\,g_e\mu_B}{2c}
\sqrt{\dfrac{\hbar\omega_c}{2}}
\vspace{0.2cm}
\\
\dfrac{\mathrm{i}\,g_0\,g_e\mu_B}{2c}
\sqrt{\dfrac{\hbar\omega_c}{2}}
&
-
\dfrac{g_e\mu_B}{2}
B_z
+
\hbar\omega_c
\end{pmatrix}
\quad,
\end{align}
for states $\ket{\downarrow,0_y},\ket{\downarrow,1_y}$, respectively. For $\underline{\underline{H}}_\uparrow$, we obtain eigenvalues
\begin{align}
\varepsilon^\uparrow_\mp
=
\dfrac{
g_e\mu_B B_z
+
\hbar\omega_c}{2}
\mp
\dfrac{\Delta_\uparrow}{2}
\quad,
\end{align}
with excitation energy independent of the static B-field
\begin{align}
\Delta_\uparrow
&=
\sqrt{
(\hbar\omega_c)^2
+
\dfrac{g^2_0 g^2_e\mu^2_B}{c}
\dfrac{\hbar\omega_c}{2}
}
\quad.
\end{align}
A Taylor expansion of $\Delta_\uparrow$ around $g_0=0$ 
\begin{align}
\Delta_\uparrow
&=
(
1
+
\dfrac{g^2_0 g^2_e\mu^2_B}{4c}
)
\hbar\omega_c
+
\mathcal{O}(g^4_0)
\quad,
\end{align}
illustrates weakly repelling eigenstates at non-zero light-matter interaction strength. For $\underline{\underline{H}}_\downarrow$, we find equivalently
\begin{align}
\varepsilon^\downarrow_\mp
&=
\dfrac{
\hbar\omega_c
-
g_e\mu_B B_z}{2}
\mp
\dfrac{\Delta_\downarrow}{2}
\quad,
\end{align}
and in particular
\begin{align}
\Delta_\downarrow
&=
\Delta_\uparrow
\quad.
\end{align}
Thus, eigenstates are always energetically well separated such that spin-polariton formation is absent for this polarization scenario.

\setcounter{equation}{0}
\renewcommand{\theequation}{\thesubsection.\arabic{equation}}
\subsection{Derivation of Cavity-Modified g-Factor}
\label{sec.derivation_cavity_g_factor}
We derive Eq.\eqref{eq.cavity_modified_g_factor} by inserting Eq.\eqref{eq.res_B_field} into Eq.\eqref{eq.cavity_zeeman_split_res} 
\begin{align}
\tilde{\Delta}_\mathrm{Zee}
&=
g_e
\mu_B
B^\star_z
-
\dfrac{g_0 g_e\mu_B}{2c}
\sqrt{
\dfrac{\hbar\omega_c}{2}
}
\quad,
\label{eq.derivation_cavity_modified_g_factor}
\vspace{0.2cm}
\\
&=
g_e
\left(
\mu_B
B^\star_z
-
\dfrac{g_0\mu_B}{2c}
\sqrt{
\dfrac{\hbar\omega_c}{2}
}
\right)
\quad,
\end{align}
and subsequently find the desired result
\begin{align}
\tilde{g}_e
&=
\dfrac{\tilde{\Delta}_\mathrm{Zee}}
{
\mu_B
B^\star_z
-
\dfrac{g_0\mu_B}{2c}
\sqrt{
\dfrac{\hbar\omega_c}{2}
}	
}
\quad,
\vspace{0.2cm}
\\
&=
\dfrac{\tilde{\Delta}_\mathrm{Zee}}{\mu_B B^\star_z}
\left(
\dfrac{1}{1-\dfrac{g_0}{2B^\star_z}\sqrt{\dfrac{\hbar\omega_c}{2}}}
\right)
\quad.
\end{align}


%


\begin{thebibliography}{200}



\bibitem{ebbesen2016} Ebbesen, T.W. Hybrid light-matter states in a molecular and material science perspective. \textit{Acc. Chem. Res.} \textbf{49}, 2403 (2016).

\bibitem{garciavidal2021} Garcia-Vidal, F.J.; Ciuti, C.; Ebbesen, T.W. Manipulating matter by strong coupling to vacuum fields. \textit{Science} \textbf{373}, 6551 (2021).

\bibitem{hutchison2012} Hutchison, J.A.; Schwartz, T.; Genet, C.; Devaux, E.;  Ebbesen, T.W. Modifying Chemical Landscapes by Coupling to Vacuum Fields. \textit{Angew. Chem. Int. Ed.} \textbf{51}, 1592 (2012).

\bibitem{thomas2016} Thomas, A.; George, J.; Shalabney, A.; Dryzhakov, M.; Varma, S.J.; Moran, J.; Chervy, T.; Zhong, X.; Devaux, E.; Genet, C.; Hutchison, J.A.; Ebbesen, T.W. Ground-state chemical reactivity under vibrational coupling to the vacuum electromagnetic field. \textit{Angew. Chem. Int. Ed.} \textbf{55}, 11462 (2016).



\bibitem{eddins2014} Eddins, A. W.; Beedle, C. C.; Hendrickson, D. N.; Friedman, J. R. Collective Coupling of a Macroscopic Number of Single-Molecule Magnets with a Microwave Cavity Mode. 
\textit{Phys. Rev. Lett.} \textbf{112}, 120501 (2014).

\bibitem{jenkins2016} Jenkins, M. D.; Zueco, D.; Roubeau, O.; Aromí, G.; Majer, J.; Luis, F. A scalable architecture for quantum computation with molecular nanomagnets. \textit{Dalton Trans.} \textbf{45}, 16682 (2016).

\bibitem{bonizzoni2017} Bonizzoni, C.; Ghirri, A.; Atzori, M.; Sorace, L.; Sessoli, R.; Affronte, M. Coherent coupling between Vanadyl Phthalocyanine spin ensemble and microwave photons: Towards integration of molecular spin qubits into quantum circuits. \textit{Sci. Rep.} \textbf{7}, 13096 (2017).


\bibitem{patrahau2024} Patrahau, B.; Piejko, M.; Mayer, R.J.; Antheaume, C.; Sangchai, T.; Ragazzon, G.; Jayachandran, A.; Devaux, E.; Genet, C.; Moran, J.; Ebbesen, T.W. Direct Observation of Polaritonic Chemistry by Nuclear Magnetic Resonance Spectroscopy. \textit{Angew. Chem. Int. Ed.} \textbf{63}, e202401368  (2024).



\bibitem{haugland2020} Haugland, T.; Ronca, E.; Kj\o nstad, E.; Rubio, A.; Koch, H. Coupled cluster theory for molecular polaritons: Changing ground and excited states. \textit{Phys. Rev. X} \textbf{10}, 041043 (2020).

\bibitem{riso2022} Riso, R.R.; Haugland, T.S.; Ronca, E.; Koch, H. Molecular orbital theory in cavity QED environments. \textit{Nat. Commun.} \textbf{13}, 1368 (2022).

\bibitem{tokatly2013} Tokatly, I.V. Time-dependent density functional theory for many-electron systems interacting with cavity photons. \textit{Phys. Rev. Lett.} \textbf{110}, 233001 (2013).

\bibitem{ruggenthaler2014} Ruggenthaler, M.; Flick, J.; Pellegrini, C.; Appel, H.; Tokatly, I.V.; Rubio, A. Quantum-electrodynamical density-functional theory: Bridging quantum optics and electronic structure theory. \textit{Phys. Rev. A} \textbf{90}, 012508 (2014).

\bibitem{schnappinger2023} Schnappinger, T.; Sidler, D.; Ruggenthaler, M.; Rubio, A., Kowalewski, M. Cavity Born-Oppenheimer Hartree-Fock Ansatz: Light-Matter Properties of Strongly Coupled Molecular Ensembles. \textit{J. Phys. Chem. Lett.} \textbf{14}, 8024 (2023).

\bibitem{fischer2024} Fischer, E.W. Cavity-modified local and non-local electronic interactions in molecular ensembles under vibrational strong coupling. \textit{J. Chem. Phys.} \textbf{161}, 164112 (2024).

\bibitem{fischer2025} Fischer, E.W. Cavity Born-Oppenheimer Coupled Cluster Theory: Towards Electron Correlation in the Vibrational Strong Light-Matter Coupling Regime. arXiv preprint arXiv:2507.10100 (2025).


\bibitem{rokaj2019} Rokaj, V.; Penz, M.; Sentef, M. A.; Ruggenthaler, M.; Rubio, A. Quantum Electrodynamical Bloch Theory with Homogeneous Magnetic Fields. \textit{Phys. Rev. Lett.} \textbf{123}, 047202 (2019).

\bibitem{rokaj2022} Rokaj, V.; Penz, M.; Sentef, M. A.; Ruggenthaler, M.; Rubio, A. Polaritonic Hofstadter butterfly and cavity control of the quantized Hall conductance. \textit{Phys. Rev. B} \textbf{105}, 205424 (2022).

\bibitem{barlini2024} Barlini, A.; Bianchi, A.; Ronca, E.; Koch, H. Theory of Magnetic Properties in Quantum Electrodynamics Environments: Application to Molecular Aromaticity. \textit{J. Chem. Theory Comput.} \textbf{20}, 7841 (2024).


\bibitem{mcweeny1965} McWeeny, R. On the Origin of Spin-Hamiltonian Parameters. \textit{J. Chem. Phys.} \textbf{42}, 1717 (1965). 

\bibitem{neese1998} Neese, F.; Solomon, E.I. Calculation of Zero-Field Splittings, g-Values, and the Relativistic Nephelauxetic Effect in Transition Metal Complexes. Application to High-Spin Ferric Complexes. \textit{Inorg.Chem.} \textbf{37}, 6568 (1998).

\bibitem{neese2017} Neese, F. Quantum Chemistry and EPR Parameters. In Goldfarb D.; Stoll, S. EPR Spectroscopy - Fundamentals and Methods. John Wiley \& Sons Ltd (2018).

\bibitem{flick2017} Flick, J.; Ruggenthaler, M.; Appel, H.; Rubio, A. Atoms and molecules in cavities, from weak to strong coupling in quantum-electrodynamics (QED) chemistry. \textit{PNAS} \textbf{114}, 3026 (2017).

\bibitem{flick2017cbo} Flick, J.; Appel, H.; Ruggenthaler, M.; Rubio, A. Cavity Born-Oppenheimer approximation for correlated electron-nuclear-photon systems. \textit{J. Chem. Theory Comput.} \textbf{13}, 1616, (2017).

\bibitem{fischer2023} Fischer, E.W.; Saalfrank, P. Beyond Cavity Born-Oppenheimer: On Nonadiabatic Coupling and Effective Ground State Hamiltonians in Vibro-Polaritonic Chemistry. \textit{J. Chem. Theory Comput.} \textbf{19}, 7215 (2023).

\end{thebibliography}
\end{document}